# A Full Frequency Masking Vocoder for Legal Eavesdropping Conversation Recording

R. F. B. Sotero Filho, H. M. de Oliveira, R. Campello de Souza

Signal Processing Group, Federal University of Pernambuco - UFPE
E-mail: rsotero@hotmail.com.br, {hmo,ricardo}@ufpe.br

*Abstract*: *This paper presents a new approach for a vocoder design based on full frequency masking by octaves in addition to a technique for spectral filling via beta probability distribution. Some psycho-acoustic characteristics of human hearing - inaudibility masking in frequency and phase - are used as a basis for the proposed algorithm. The results confirm that this technique may be useful to save bandwidth in applications requiring intelligibility. It is recommended for the legal eavesdropping of long voice conversations.*

The purpose of the voice compression is to obtain a concise representation of the signal, which allows efficient storage and transmission of voice data [1]. With proper processing, a voice signal can be analyzed and encoded at low data rates and then resynthesized. In many applications, the digital coding of voice is needed to introduce encryption algorithms (for security) or error correction techniques (to mitigate the noise of the transmission channel). Often, the available bandwidth for the transmission of digitized voice is a few kilohertz [2]. In such conditions of scarce bandwidth, it is necessary to adopt coding schemes that reduce the bit rate in such a way that information can be properly transmitted. However, these coding systems at low bit rate cannot reproduce the speech waveform in its original format. Instead, a set of parameters are extracted from the voice, transmitted and used to generate a new waveform at the receiver. This waveform may not necessarily recreate the original waveform in appearance, but it should be perceptually similar to it [3]. This type of encoder – called the vocoder (a contraction from voice encoder), a term also used broadly to refer to encoding analysis / synthesis in general, will use perceptually relevant features of the voice signal to represent it in a more efficient way, without compromising much on its quality [3]. The vocoder was first described by Homer Dudley at Bell Telephone Laboratory in 1939, and consisted of a voice synthesizer operated manually [4]. Generally speaking, the vocoders are based on the fact that the vocal tract changes slowly and its state and configuration may be represented by a set of parameters. Typically, these parameters are extracted from the spectrum of the voice signal and updated every 10-25 ms [5]. In general, given its low complexity in the process of generating the synthesized voice, the modeling, and the nature of simplifications carried out by vocoders, they introduce losses and/or distortions that ultimately make the voice quality below those obtained by waveform encoders [5]. Two properties of voice communication are heavily exploited by vocoders. The first is the limitation of the human auditory system [6]. This restriction makes the listeners hearing rather insensitive to various flaws in the process of voice reproduction. The second concerns the physiology of the voice generation process that places strong constraints on the type of signal that can occur, and this fact can be exploited to model some aspects of the production of the human voice [3,5]. The vocoder also find wide acceptance as an essential principle for handling audio files. For example, audio effects like time stretching or pitch transposition are easily achieved by a vocoder [7]. Since then, a series of modifications and improvements to this technology have been published [5]. In this article we present an innovative technique, which combines simplicity of implementation, low computational complexity, low bit rate and acceptable quality of generated voice files. In our approach, the stage of analysis of the voice signal is based on full frequency masking, recently published in [8] and explained in detail in Section III. In the resynthesis stage of the signal, we present a new approach based on spectral filling by a beta probability distribution.

The first stage of the proposed vocoder is a pre-signal processing. This is often required in speech processing, since the characteristics of voice signals have peculiarities that need to be worked with, previously. Because vocoders are designed for voice signals, which have most of their energy concentrated in a limited range of frequencies (typically between 300 Hz and less than 4 kHz), it is required to limit the bandwidth of the signals within this range, by a low-pass filter. Then a sampling rate that meets the Shannon sampling theorem condition must be taken. According to this theorem [9], there is no loss of information in the sampling process when a signal band limited to $f_m$ Hz is sampled at a rate of at least $2f_m$ equally spaced samples per second. *Voice Segmentation and Windowing*-A signal is said to be stationary when its statistical features do not vary with time [9]. Since the voice signal is an stochastic process, and knowing that the vocal tract changes its shape very slowly in a continuous speech, many parts of the acoustic waveform can be assumed as stationary within a short duration range (typically between 10 and 40 ms). Segmentation is the partition of the speech signal into pieces (frames), selected by windows of duration perfectly defined. The size of these segments is chosen within the bounds of stationarity of the signal [10]. The use of windowing is a way of achieving increased spectral information from a sampled signal [11]. This "increase" of information is due to the minimization of the margins of transition in truncated waveforms and a better separation of the signal of small amplitude from a signal of high amplitude with frequencies very close to each other. Many different types of windows can be used. The Hamming window was chosen due to the fact that it presents interesting spectral characteristics and softness at the edges [12]. *Pre-emphasis-* The pre-emphasis aims to reduce a spectral inclination of approximately -6dB/octave, radiated from the lips during speech. This spectral distortion can be eliminated by ap-plying a filter response approximately +6 dB / octave, which causes a flattening of the spectrum [13]. The hearing is less sensitive to frequencies above 1 kHz of the spectrum; pre-emphasis amplifies this area of the spectrum, helping spectral analysis algorithms for modeling the perceptually aspects of the spectrum of voice [6,11]. Equation (1) describes the pre-emphasis performed on the signal that is obtained by differentiating the input.

$$y(n)= x(n)-a.x(n-1), \qquad (1)$$

for $1 \leq n < M$, where $M$ is the number of samples of $x(n)$, $y(n)$ is the emphasized signal and the constant "$a$" is normally set between 0.9 and 1. In this paper the adopted value was "$a$" equals to 0.95 [13]. The algorithms developed for implementation of this vocoder were written in MATLAB$^{TM}$ platform, owing to the fact that it is a widespread language in the academic world and it is easy to implement. In the following, details of the approach are described. As in most efficient speech coding systems, vocoders may exploit certain properties of the human auditory system, taking advantage of them to reduce the bit rate. The technique proposed in this article for implementation of the vocoder is founded on two important characteristics: the masking in frequency and the insensitivity to phase. The function of the stage of analysis is, a priori, to identify the frequency masking in the spectrum of the signal (obtained by an FFT of blocklength 160), partitioned into octave bands, discard signals that "would not be audible," due to the phenomenon of masking in frequency [14], and totally disregard the signal phase.

*Psycho-Acoustics of the Human Auditory System-* Because it is of great importance for the understanding of the proposed method, a few characteristics of human auditory system are briefly discussed [6,14].

- Frequency Masking: Masking in frequency or "reduced audibility of a sound due to the presence of another" is one of the main psycho-acoustic characteristics of human hearing. The auditory masking (which may be in frequency or in time) occurs when a sound, that could be heard, is often masked by another, more intense, which is in a nearby frequency. In general, the presence of a tone cannot be detected if the power of the noise is more than a few dB above this tone. Due to the effect of masking, the human auditory system is not sensitive to detailed structure of the spectrum of a sound within this band [3,5].

- Insensitivity to the phase: The human ear has little sensitivity to the phase of signals. The process can be explained by examining how sound propagates in an environment.

Any sound that propagates reaches our ears through various obstacles and travels distinct paths. Part of the sound gets lagged, but this difference is hardly felt by the ear [15]. The information in the human voice is mostly concentrated in "bands of frequencies". Based on this fact, the proposed vocoder discards the phase characteristics of the spectrum.

*Simplification of the spectrum via the frequency masking-* Equipped with the pre-processed signals, we can start the stage of signal analysis, which is described in the sequel. For each voice segment of the file, an FFT of blocklength 160 (number of samples contained in a frame of 20 ms of voice) is applied, thus obtaining the spectral representation of each voice frame. Only the magnitude of the spectrum is considered. After that, the spectrum is segmented into regions of influence (octaves). The range of frequencies between 32 and 64 Hz is removed from the analysis. The first pertinent octave corresponds to the frequency range 64 Hz-128 Hz, the second covering the band 128 Hz-512 Hz, and so on. The sixth (last octave band) matches the range of 2048 Hz-4000 Hz (remarking that from here the spectrum produced by the FFT begins to repeat). Since the sampling rate is 8 kHz, each spectral sample corresponds to a multiple of 50 Hz, and the first sample represents the DC component of each frame of speech. Because this sample has no information, it is promptly disregarded from the analysis. Since the spectral lines have a step of 50 Hz, the first octave (from 64 Hz to 128 Hz) is represented by the spectral sample of 100 Hz, the second octave (from 128 Hz to 256 Hz) by samples at 150 Hz, 200 Hz and 250 Hz, with the remaining octaves following a similar reasoning. After this preliminary procedure, we search at each octave, in all relevant sub-bands of the voice signal, for the DFT component of greatest magnitude, i.e., that one that (potentially) can mask the others. There are 80 spectral lines (dc is not shown). This component is taken as the sole representative tone in each octave (as an option of reducing the complexity). The other spectral lines are discarded, assuming a zero spectral value. A total of 79 frequencies coming from the estimation of the DFT with $N$=160 is then reduced to only 4 survivors (holding less than 5% of the spectral components). Therefore, each frame is now represented in the frequency domain by 4 pure (masking) tones. This technique is called full frequency masking [8]. These simplified frames are encoded and used by a synthesizer to retrieve the voice signal. Now a signal synthesis is described on the basis of a spectral filling via a distribution of probability. The beta distribution is a continuous probability distribution defined over the interval $0 \leq x \leq 1$, characterized by a pair of parameters $\alpha$ and $\beta$, according to Equation [16]:

$$P(x)=1/B(\alpha,\beta)\ x^{(\alpha-1)} (1-x)^{(\beta-1)}, \qquad 1<\alpha,\beta<+\infty, \qquad (2)$$

whose normalized factor is $B(\alpha,\beta)=(\Gamma(\alpha)\Gamma(\beta))/(\Gamma(\alpha+\beta))$, where $\Gamma(.)$ is the generalized Euler factorial function and $B(.,.)$ is the Beta function. The point where the maximum of the density is achieved is the mode and can be computed by the following equation [16]:

$$mode= (\alpha-1)/(\alpha+\beta-2). \qquad (3)$$

| Octave (Hz) | # spectral samples/octave |
|---|---|
| 32-64 | 1 |
| 64-128 | 1 |
| 128-256 | 3 |
| 256-512 | 5 |
| 512-1024 | 10 |
| 1024-2048 | 20 |
| 2048-4096 | 39 |

Table 1. Number of Spectral Lines per Octave Estimated by a DFT of Length $N$=160 with a Sample Rate 8 kHz.

The purpose of the synthesis stage is to retrieve the voice signal from data provided by the parsing stage. As mentioned, the full frequency masking was adopted to simplify the spectrum of each frame of voice. Such a simplification results in a very vague and spaced sample configuration in the spectrum. To improve this representation, the synthesizer can use the spectral filling technique via beta distribution, so as to smooth the abrupt transition between adjacent samples in octaves, assigning interpolated values to lines with zero magnitude, thus filling up the spectrum completely. Each octave has its own distribution and these are updated with each new frame. The peak of each of these distributions is equal to the survivor spectral sample after the full masking simplification. In what follows, the methodology of spectral filling, via beta distribution, is described. Since the beta distribution is defined over the interval [0,1], see Fig.1, it is necessary to scale and translate the original expression of the distribution, so that their range encompass the transition from one octave to another. Moreover, the value of the mode should assume the same value of the survivor spectral sample within the octave. Based on the original expression of the beta distribution, given by Eq. (2), there is a suitable scaling of the curve so that the upper limit is equivalent to the difference between the normalized cutoff frequency exceeding ($f_M$) and lower ($f_m$) of each octave, i.e., $f_M - f_m$. The cutoff frequencies need to be normalized, since the limiting frequency of octaves (64-128 Hz, 128 – 256 Hz, etc.) are not multiples of 50 Hz, which is the value of the spectrum step while sampling at 8 kHz. Later, the curve must be translated so that the lower and upper limits become $f_m$ and $f_M$, respectively. By making the fitting, it is also necessary to adjust the value of the mode, which becomes

$$new_{mode}= (\alpha-1)/(\alpha+\beta-2) (f_M - f_m)+f_m. \quad (4)$$

From this expression and after some mathematical manipulations, we find a relation between $\alpha$ and $\beta$, which is useful in representing the adjusted expression of the distribution:

$$\beta-1=(\alpha-1).Q, \quad (5)$$

where:

$$Q:= (f_M - f_c)/(f_c - f_m). \quad (6)$$

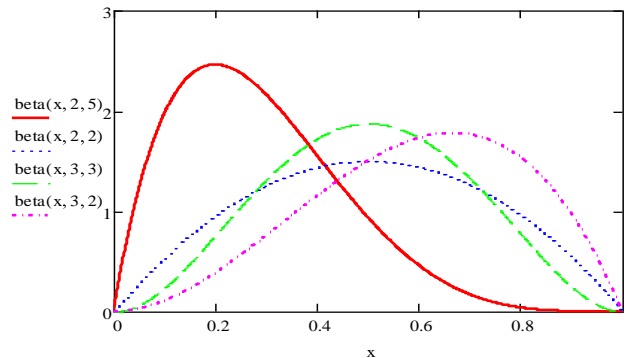

Figure 1. Envelope shape of the survivor tone is shown for a few parameters $\alpha$ and $\beta$.

The final expression, one that is used to fulfill the spectral algorithm each frame, is given by:

$$P(x)= 1/(f_M - f_m)^{(\alpha+\beta-2)} (x - f_m)^{(\alpha-1)} (f_M - x)^{(\beta-1)}. \quad (7)$$

The value of $\alpha$ in Eq. (7) represents a parameter of expansion/compression of the interpolation curve. The higher its value, the narrower it becomes. The values of $\alpha$ were octave-dependent. Fig. 2 shows the magnitude of the spectrum of a frame of a file (test voice file), (a) before simplifying by masking, (d) after simplifying and (c) after the fulfilling via beta distribution. A few audio files generated by this vocoder are available at the URL http://www2.ee.ufpe.br/codec/vocoder.html Given the symmetry of the DFT, it is also necessary to fulfill the half mirror portion of the spectrum for proper signal restoration. Otherwise a signal in time domain, complex in nature, will be incorrectly generated. As one of the last stages of reconstruction of the voice signal, there is the transformation from the frequency domain to the time domain of all voice frames. Such a transformation is achieved through the inverse fast Fourier transform (IFFT) of the same blocklength of a frame. In doing so, the frames are glued one by one, resetting the pre-emphasized signal. An inverse pre-emphasis filter is used to de-emphasize the signal, thus finalizing the process of recovering the voice signal. For each frame, spectral samples survivors and the positions of each are then quantized and encoded, and saved in a binary format (.voz) and used later by a synthesizer.

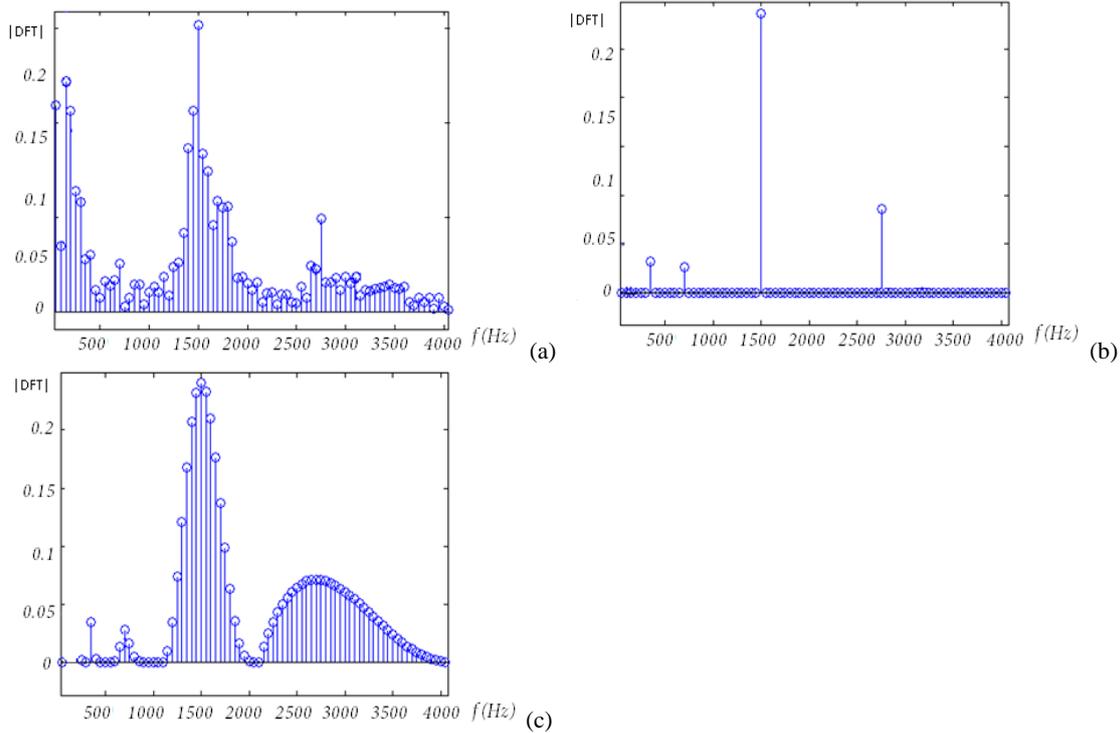

Figure 2. Steps of the procedure of analysis/synthesis of a frame of tested voice signal. The spectrum of a voice frame computed by the FFT is shown: a) Original spectrum, b) Simplified spectrum using full masking, c) Spectrum fulfilled by the beta distribution.

The quantization and coding procedures (allocation of bits per frame) are shown in the sequel. The most common method was used in the quantization of frames: the uniform quantization. A number of levels coincident with a power of 2 was adopted to simplify the binary encoding. The maximum excursion of the signal (greater magnitude of the full spectrum of the voice signal) was thus divided into 256 intervals of equal length, each represented by one byte. Since there are no negative samples to be quantized (the magnitude of the spectrum does not assume negative values), the quantizer cannot be bipolar.

| Relevant octave | #possible survivor components | Bits A+P |
| --- | --- | --- |
| #1 (256-512 Hz) | 5 | 8 + 3 |
| #2 (512-1024 Hz) | 10 | 8 + 4 |
| #3 (1024-2048 Hz) | 20 | 8 + 5 |
| #4 (2048-4096 Hz) | 39 | 8 + 6 |

Table 2. Bit allocation in a voice frame (20 ms). The required number of bits is expressed as A + P, where A is the number of bits for spectral line amplitude and P the number of bits to express the relative position within the OCTAVE.

A MATLAB routine is specifically designed for this purpose. The quantization of the positions was not necessary, since they are integer-valued. In order to reduce the number of bits needed for encoding voice frames, the bit allocation algorithm took into consideration the bandwidth of each octave. A lower octave reduces by half the bandwidth and therefore fewer bits are needed for proper co-ing of positions in which the spectral masking occurred. Positions in successive octaves (spectrum towards high frequencies) need an extra bit for its correct representation. For example, a tone masking which occurs in the first octave (256-512 Hz) has 5 possible occurrences (position 7 to position 11 of DFT), thereby requiring a 3-bit codeword. In the next octave, (512 - 1024 Hz), the maximum position that the tone masking may occur is 21, which can be encoded by a 4-bit codeword. In the subsequent two octaves, the peak may be at 41th position (5-bit codeword) and 80th (6-bit codeword), respectively. For the maximum values of the spectral masking samples, one byte is reserved for their representation. The number of bits allocated to each of these parameters is shown in Table 2. As mentioned, the phase information of the spectrum is disregarded. It is seen that each voice frame needs only 50 bits (18 for identifying positions and 32 for identifying masking tones), leading to a rate of 50 bits/20 ms=2.5 kbps. *The binary format .voz* - The bit allocation in each frame, summarized in Table II, suggests the concatenation of encoded frames. The representation of a voice frame in this format (extension .voz) is shown in Fig.3. The 50 bits are distributed into four sub blocks (one for each octave), indicating the value of the spectral sample followed by its respective position in the spectrum. The voice files registered in the .wav format are all converted to this binary format, by a Matlab routine. In the decoder, the reconstruction algorithm of the synthesized spectrum, can recover the voice signal by converting it back into the .wav format.

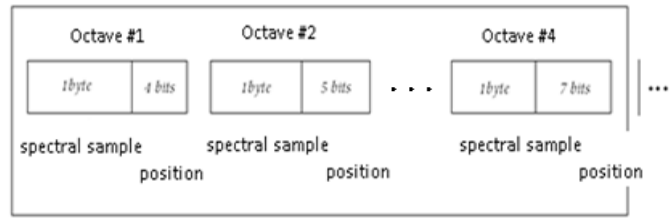

Figure 3. Frame of files in the format .voz (20 ms).

Simulation results usually focus on intelligibility and voice quality versus bit rate [17]. Fifty-eight subjects, of whom eight were trained, were accessed for this study. Voice quality is estimated using the "Mean Opinion Score (MOS)" and "degradation Mean Opinion Score (DMOS)" tests. During the tests, "MOS"-listeners were asked to rate voice quality of the output files considering an absolute scale 1-5, with 1 meaning very poor quality and 5 being excellent. The main obstacle for the MOS testing was that ordinary people were not familiar with low bit rate vocoders and got confused between a sound disharmonies, stuffy, with tinnitus, and the nasal quality of speech and noise added after encoding. To overcome this limitation, DMOS tests were conducted. In this test, listeners were asked to rate the quality of sentences encoded and spread over time on the output of the vocoder MELP pattern [15]. Preliminary tests were conducted and voice signals tested using four different techniques of synthesis.

*Evaluated techniques.*
1. Synthesized signals with no spectral filling.
2. Vocoder signals reconstructed via beta spectral filling technique.
3. Synthesized voice signals combining 1 and 2 techniques (linear combination).
4. Voice signals from item 2, but with an extra Hamming windowing.

Results are summarized in Table III. They were reasonable, given the low bit rate (2.5 kbits/s) and low implementation complexity of the vocoder. Indeed, the comparison is "unfair" to those coders, since the MOS values obtained for them were much more insightful and performed with a wide range of listeners, or even using objective methods such as PESQ [17]. It can be observed from Table 3 that noise is still a factor that impairs such an assessment, reflecting a lower MOS score for noisy signals (produced by the technique of spectral filling).

| Vocoder technique | MOS score |
|---|---|
| 1 | 3.0 |
| 2 | 2.5 |
| 3 | 2.8 |
| 4 | 3.0 |

Table 3. MOS scores for the voice signals synthesized by four different techniques.

We introduced a new vocoder that can represent a voice signal using fewer samples of the spectrum. Our initial results suggest that this approach has the potential to transmit voice, with acceptable quality, at a rate of a few kbits/s. A new technique of spectral filling was also presented, which is based on the beta distribution of probability. Surprisingly, this was not helpful in improving the voice quality, although it improved the naturalness of the speech generated by this vocoder. This vocoder can be useful for the transmission of maintenance voice channels in large plants. It was successfully applied in a recent speaker recognition system. In particular, it is offered as a technique for monitoring long voice conversation stemming from authorized eavesdropping.

# References

[1] Schroeder, M.R., A Brief History of Synthetic Speech, Speech Comm., vol.13, pp.231-237, (1993).

[2] Pope, S.P., Solberg, B., Brodersen, R.W, A Single-Chip Linear-Predictive-Coding Vocoder, IEEE J. of Solid-State Circuits, vol. SC-22, (1987).

[3] Holmes, J., Holmes, W. Speech Synthesis and Recognition, Taylor & Francis, 2001.

[4] Schroeder, M.R., Homer Dudley: A tribute, Signal Processing, vol.3, pp.187-188, (1981).

[5] Spanias, A., Speech Coding: A tutorial Review, Proc. of IEEE, vol. 82, pp.1541-1582, (1994).

[6] Greenwood, D., Auditory Masking and the Critical Band, J. Acoust. Soc. Am., vol. 33, pp. 484-502, (1961).

[7] Zoelzer, U. Digital Audio Effects', Wiley & Sons, pp.201-298, 2002.

[8] Sotero Filho, R.F.B, de Oliveira, H., Reconhecimento de Locutor Baseado no Mascaramento Pleno em Frequência por Oitava, Audio Engineering Congress, AES2009, São Paulo,2009.

[9] Lathi, B.P. Modern Digital and Analog Communication Systems. Oxford Univ. Press, NY, 1998.

[10] Rabiner, L.R.; Schafer, R.W., Digital Processing of Speech Signals. Prentice Hall, NJ, 1978.

[11] Turk, O., Arsan, L.M., Robust Processing Techniques for Voice Conversation, Computer Speech & Language, vol. 20, pp.441-467, (2006).

[12] Taubin, G., Zhang, T., Golub, G., Optimal Surface Smoothing as Filter Design, Lecture Notes on Computer Science, vol. 1064, pp. 283-292, (1996).

[13] Schnell, K., Lacroix, A., Time-varying pre-emphasis and inverse filtering of Speech, Proc. Interspeech, Antwerp, 2007.

[14] Wegel, R.L., Lane, C.E., Auditory masking of one pure tone by another and its probable relation to the dynamics of the inner ear, Physical Review, vol. 23, pp. 266-285, (1924).

[15] Smith, S.W., Digital Signal Processing – A Practical Guide for Engineers and Scientists, Newnes, 2003.

[16] de Oliveira, H.M., Araújo G.A.A., Compactly Supported One-cyclic Wavelets Derived from Beta Distributions, Journal of Communication and Information Systems, vol. 20, pp.27-33, (2005).

[17] Kreiman, J., Gerrat, B.R., Validity of rating scale measures of voice quality, J. Acoust. Soc. Am., vol. 104, pp.1598-1608, (1998).